\documentclass[authoryear,final,3p,12pt]{elsarticle}
\usepackage{verbatim}
\usepackage{colortbl,booktabs}
\usepackage{xcolor}
\usepackage{graphicx}
\usepackage[font=small]{caption}

\usepackage{color}
\usepackage{amsmath,bm}
\usepackage{amsmath}
\usepackage{amssymb}
\usepackage{algorithm}
\usepackage{algpseudocode}
\usepackage{amsthm}
\usepackage{makecell}

\newtheorem{proposition}{Proposition}

\newtheorem{remark}{Remark}
\usepackage{unitsdef}

\usepackage[utf8]{inputenc}

\usepackage{hyperref}

\usepackage{nomencl}
\makenomenclature

\usepackage{etoolbox}
\renewcommand\nomgroup[1]{%
  \item[\bfseries
  \ifstrequal{#1}{P}{Sets}{%
  \ifstrequal{#1}{N}{Variables}{%
  \ifstrequal{#1}{O}{Constants}{}}}%
]}

\journal{Elsevier}

\begin{document}

\begin{frontmatter}

\title{Efficient Demand Response Location Targeting for  Price Spike Mitigation by Exploiting Price-demand Relationship}

% \title{Demand Response Targeting for Lowering Electricity Price: a Mixed-integer Quadratic Programming Approach}

% \author{Yufan Zhang, Honglin Wen, Tao Feng, and Yize Chen

% 	\thanks{Yufan Zhang and Honglin Wen are with the Department of Electrical Engineering, Shanghai Jiao Tong University,  emails: \{zhangyufan, linlin00\}@sjtu.edu.cn. Tao Feng is with the School of Transportation and Logistics, Southwest Jiaotong University,  email: fontoo@my.swjtu.edu.cn. Yize Chen is with the Artificial Intelligence Thrust, Hong Kong University of Science and Technology (Guangzhou), email: yizechen@ust.hk.\\
% Co-corresponding authors: Honglin Wen and Yize Chen}
%  \vspace{-3em}
% }


% \maketitle
% \thispagestyle{empty}
% \pagestyle{plain}
%\emph{arabic}

\author[inst1]{Yufan Zhang}
\author[inst2]{Honglin Wen\corref{cor1}}
\ead{linlin00@sjtu.edu.cn}
\cortext[cor1]{Co-corresponding authors}

\author[inst3]{Tao Feng}
\author[inst4]{Yize Chen\corref{cor1}}
\ead{cyz1212wj@gmail.com}

\affiliation[inst1]{organization={Department of Electrical and Computer Engineering, University of California San Diego}}

\affiliation[inst2]{organization={Department of Electrical Engineering, Shanghai Jiao Tong University}}

\affiliation[inst3]{organization={Department of Logistics and Maritime Studies, The Hong Kong Polytechnic University}}

\affiliation[inst4]{organization={Artificial Intelligence Thrust, Hong Kong University of Science and Technology (Guangzhou)}}

%%%%%%%%%%%%%%%%%%%%%%%%%%%%%%%%%%%%%%%%%%%%%%%%%%%%%%%%%%%%%%%%%%%%%%%%%%%%%%%%
\begin{abstract}
\textcolor{black}{Demand response (DR) leverages demand-side flexibility, offering a promising approach to enhance market conditions like mitigating wholesale price spikes. However, poorly chosen DR locations can inadvertently increase electricity prices. For that, we introduce a method to rigorously select DR locations and corresponding demand reductions. We formulate a bilevel program where the upper level determines the DR locations and demand reductions while ensuring the average nodal prices meet a predetermined target. The lower level tackles an economic dispatch (ED) problem and feeds the resulting nodal prices back to the upper level based on post-DR demands. This bilevel formulation presents challenges due to the lower-level non-convexity affecting the upper-level constraints on average nodal prices. To address this, we propose to replace the lower level with a piecewise linear function representing the price-demand relationship, solving iteratively for each linear segment. This results in a tractable mixed-integer linear program.} An acceleration strategy is proposed to further reduce the computation time. Numerical studies demonstrate the ability of the proposed approach to reduce prices to a desired level. Besides, we empirically show that the proposed approach is robust against inaccurate system parameters and can reduce computation time by over 50\%.

% In power networks, nodal electricity prices are derived by solving the economic dispatch (ED) problem, where the electric demands act as input parameters. However, solving the ED in a DR program can impose an unnecessary computational burden, particularly when only a specific set of demand nodes participate in DR within a large-scale system. In this paper, we derive the price-demand relationship within the ED problem using multi-parametric programming theory offline, for improving real-time efficiency of DR solving process. With such a price-demand relationship at hand, the price associated with specified demand vector can be simply obtained during the operation stage. We demonstrate the effectiveness of our proposal by designing DR programs for mitigating price spikes. The proposed approach can efficiently select a group of nodes for DR program implementation. For that, we formulate a mixed-integer quadratic program and introduce a solution strategy to expedite computation. Numerical studies demonstrate the ability of the proposed approach to reduce prices to a desired level. Besides, we empirically show that the proposed approach is robust against inaccurate system parameters and can reduce computation time by over 50\%.

\begin{keyword} Demand response, price spike, price-demand relationship, multi-parametric programming
\end{keyword}

\end{abstract}

%%%%%%%%%%%%%%%%%%%%%%%%%%%%%%%%%%%%%%%%%%%%%%%%%%%%%%%%%%%%%%%%%%%%%%%%%%%%%%%%
\end{frontmatter}

\nomenclature[P]{\(\mathcal{N}\)}{The set of nodes in the transmission network}
\nomenclature[P]{\(\mathcal{A}\)}{The set of arcs in the transmission network}
% \nomenclature[P]{\( \mathcal{L}\)}{Initial convex set of all feasible vectors of loads}
\nomenclature[N]{\(\textcolor{black}{P_i^\text{g}}\)}{Power output of the generator connected to node $i$}
\nomenclature[N]{\(\bm{\lambda}\)}{Locational marginal prices}
\nomenclature[N]{\(\bm{x}\)}{Nodal demand reductions}
% \nomenclature[N]{\(\hat{\bm{l}}\)}{Nodal demands after DR}
\nomenclature[N]{\(\bm{v}\)}{Binary vector indicating DR/non-DR locations}
\nomenclature[N]{\(\textcolor{black}{\bm{P}^\text{g}}\)}{Power outputs of all generators in the network}
\nomenclature[O]{\(\textcolor{black}{a_i^\text{g},b_i^\text{g},c_i^\text{g}}\)}{Coefficients of the quadratic generation cost}
\nomenclature[O]{\(\ \bm{l}\)}{Nodal demands before DR}
\nomenclature[O]{\(\ \bm{\tau}\)}{Cost for per unit demand reduction}
\nomenclature[O]{\(\ K\)}{The maximum number of DR targeted nodes}
\nomenclature[O]{\(\ \hat{\lambda}\)}{Reference value of average locational marginal price}
\nomenclature[O]{\(\ \textcolor{black}{\bm{P}^{\text{min}},\bm{P}^{\text{max}}}\)}{Minimum and maximum output power of generators}
\nomenclature[O]{\(\ \bm{H}\)}{Shift factor matrix}
\nomenclature[O]{\(\ \bm{f}\)}{Maximum transmission capacity}
\printnomenclature

\section{Introduction}
\subsection{Background and Motivation}

Extreme weather events, insufficient infrastructure investments, and unmet electricity demand could all lead to skyrocketing and volatile electricity prices in the wholesale market \citep{doering2021effects,mulhall2014energy}. For instance, in July 2022, the average wholesale price reaches 182 $\$/MWh$ in the electricity markets of Texas, due to the record-breaking electricity demand \citep{News}. Such price spikes bring disruptive consequences and serious challenges to market operations, as abrupt electricity price changes are unfavorable for electricity market and power grid operations \citep{zareipour2007electricity}. Moreover, under the conditions of high demand or tight supply, it is non-trivial to identify whether high prices result from market forces abuse or imperfect competition, which can impair the market efficiency \citep{meritet2014market}. Once the causes of price spike are  anti-competitive practices or market failure to match supply and demand, measures for mitigating price spikes are required to be taken \citep{sirin2023market}.  Therefore, efforts have been directed towards either identifying the causes of price spikes \citep{lin2022spatio,doering2021effects,su2020compound} or mitigating them \citep{mari2014hedging}. \textcolor{black}{Over the past decade, demand-side management and in particular demand response (DR) offers the great potential to help mitigate the price spike \citep{yuan2016distribution,morais2014demand,nolan2015challenges,asadinejad2017optimal}.} This is made possible as the shifted electricity demand can result in generation cost reduction by changing operation patterns (such as the change of marginal generators or binding lines), and in turn, a reduction in electricity price.

Indeed, the nodal electricity prices are commonly spatially differentiated and referred to as the locational marginal prices (LMPs). Considering that the LMP at each node represents an additional cost incurred when adding one more unit of nodal demand, it seems logical to prioritize demand reduction at nodes with relatively high LMPs to mitigate price spikes. However, previous studies have shown that the heuristic approach does not always yield the desirable results \citep{LEE2022103723}. \textcolor{black}{Another intuition is to implement demand response (DR) at the nodes most sensitive to operational requirements, such as the need for reserves \citep{wang2011event} or voltage satisfaction \citep{davarzani2019implementation}. However, such an approach limits the demand reduction to a small value so that the sensitivities can be considered constant, which does not fully utilize the DR capacity. Also, there is no guarantee that the selected nodes are optimal.} If DR programs are improperly designed in terms of locations, adverse effects including higher price spikes can happen~\citep{ wu2013impact,yang2009demand}, which is an improper use of demand-side resources.  Network congestion is one of the reasons behind the higher price spikes \citep{stawska2021demand}. For instance, even if the demands reduce, the generation from the cheap-to-run power plants cannot be fully utilized, due to the limited transmission lines capacity. 

Therefore, how to determine the eligible set of locations to implement DR is essential. Following \citep{LEE2022103723}, we refer to this task as \emph{DR targeting} in this work. This task is not only crucial for employing DR to mitigate price spikes \citep{aazami2011demand} but also essential for leveraging DR to alleviate transmission line congestion \citep{dehnavi2016determining}.  However, addressing such a task is not trivial, as the relationship between LMPs and demands is implicitly embedded within an economic dispatch (ED) problem. For instance, previous studies leverage heuristic approach \citep{LEE2022103723}, which takes a reference of historical load data in the group where associated average LMP is below the threshold. Then it determines nodes for implementing DR as well as the nodal demand reduction quantity to let the adjusted load track the selected reference. However, such an approach \citep{LEE2022103723} has not exploited the structure of the underlying ED problem and operational conditions, thereby the resulting DR decisions may not be optimal. This motivates the exploration of the following technical
question: \textit{Is it possible to explicitly and theoretically derive the price-demand relationship to help rigorously identify locations of implementing DR program for mitigating price spike?} In this work, we address such a challenge by multiparametric programming theory~\citep{tondel2003algorithm,Grancharova2012}.

\subsection{Related Works}

%\subsubsection{Price Spike Mitigation via DR}

\textit{Price Spike Mitigation via DR:} In electricity markets and system operations, DR can play a pivotal role to mitigate price spike and reduce system costs. Usually, entities such as load serving entities (LSE) design price- or incentive- based mechanisms to collect flexible resources from end users \citep{asadinejad2017optimal}. Then, those distribution-level entities offer flexibility to transmission system operator (TSO) for reducing the price spike. We group the studies leveraging DR  for mitigating price spike into two categories. The first category considers distribution-level entities as leaders, which leverage flexibility offered by DR in response to high electricity price. A bilevel program is formulated, where strategic entities are at the upper level designing power injection at the interface of TSO, to maximize thier profits \citep{zhong2012coupon,han2022optimal} or to minimize the peak load and load fluctuation \citep{park2021optimal}. In the solution strategy of such a bilevel program, the Karush–Kuhn–Tucker (KKT) conditions of the ED are leveraged to transform it into a single-level one \citep{nizami2020residential,de2020optimal}. The second category considers TSO as leaders, which solves an ED problem and incorporates the elastic load consumed by distribution-level entities \citep{aazami2011demand}. For instance, \citep{ding2020tracking} propose an iterative approach and obtain the equilibrium of LMP after implementing DR. Our work falls in the second category. However, we not only address the demand consumption after implementing DR, but also identify the proper DR locations.

\textit{Price-demand Relationship  Embedded within ED:} The relationship between the nodal LMPs and the nodal demands is implicitly reflected in the ED problem.  Previous studies \citep{zhou2011short} showed that the 
nodal LMPs relates to a distinct ED operation pattern, which is associated with a set of binding constraints. Such binding constraints remain unchanged, when the demands vary in a \emph{a unique region in demand space}. For instance, for an ED problem with the linear objective, the nodal LMPs remain unchanged when demands lie in  a particular region with the same set of binding ED constraints.  Though some of  literature leverage the heuristics in such a relationship to transform the LMP prediction task to a region classification for both probabilistic~\citep{ji2016probabilistic} and deterministic~\citep{radovanovic2019holistic} forecasts, how to leverage it for mitigating price spikes is unclear.

\subsection{Proposed Method and Contribution}

In this paper, we propose a novel formulation and effective solution approach for mitigating the price spike via DR targeting. Concretely, a bilevel mixed-integer program is formulated, where the upper-level determines the DR locations (indicated by binary variables) and demand reduction, to minimize the total cost of implementing DR while ensuring the average LMP reduces to a reference value. The lower-level solves the ED problem with the demand after DR as input, and returns the LMPs to the upper level. In this way, the upper level ``controls'' LMPs via determining demands. Solving such a bilevel program is not trivial, as the relationship between the LMPs and demands embedded in the lower-level ED is not necessarily convex. This makes the upper-level constraints regarding average LMP become a nonconvex one. We attempt to address this problem by explicitly deriving the price-demand relationship, which is ensured by multiparametric theory to be a piecewise linear function for the ED modeled by a quadratic program~\citep{tondel2003algorithm,Grancharova2012}. We propose to replace the lower-level program with each linear piece of the piecewise linear function and reduce the bilevel program to a single-level one, which is a tractable mixed integer linear program (MILP). We solve such an MILP over each piece, and find the one with the minimum optimal demand reduction cost. To further reduce the computation time, we propose to first judge the feasibility of the relaxed problem of MILP, which is a linear program (LP),  before solving the MILP problem.

The main contributions of this paper can be summarized as follows:

\textcolor{black}{1) A solution strategy to the DR targeting problem for mitigating price spike, which rigorously determines \emph{both locations and demand reductions for DR}.}

% 1) A solution strategy to the DR targeting problem for mitigating price spike, which transforms the complex problem to a set of MILPs to rigorously determine \emph{both the locations and demand reductions for DR}.
% With the prior obtained price-demand relationship at hand, the need of resorting to ED problem is spared, which improves the computational efficiency.

2) A theoretical derivation of the price-demand relationship between nodal LMPs and nodal demands, which replaces the lower level in bilevel DR targeting problem.

3) An acceleration strategy to reduce the computation time, which significantly reduces the computation burden for the original problem and makes the proposed framework implementable in a real market environment. 

In addition, we also demonstrate the proposed framework can be readily applied to a variant of settings, such as under inaccurate network models. The remaining sections of this paper are organized as follows. Section 2 introduces the preliminaries of ED, whereas Section 3 presents the physical framework and formulates the problem. The details of price-demand relationship derivation and solution strategy are given in  Section 4 and 5, respectively. Results are discussed and evaluated in Section 6, followed by the conclusions.

\section{Preliminaries: Economic Dispatch}

In this paper, we consider a transmission network represented by the graph $\mathcal{G}(\mathcal{N},\mathcal{A})$, where $\mathcal{N}$ is the set of nodes and $\mathcal{A}$ is the set of lines. The ED problem considers the minimization of the total generation cost $\sum_{i=1}^{|\mathcal{N}|}C_\text{g}(P_i^\text{g})$, where $P_i^\text{g}$ is the power output of the generator connected to the node $i$, and $C_\text{g}(\cdot)$ is the generation cost function. Without loss of generality, the quadratic cost is considered in our framework, i.e., $C_\text{g}(P_i^\text{g})=\frac{1}{2}a_i^\text{g}\cdot(P_i^\text{g})^2+b_i^\text{g}\cdot P_i^\text{g}+c_i^\text{g}$. \textcolor{black}{The cost function is contingent upon the market rules governing the format of offers. Most electricity markets use linear or quadratic cost functions \citep{kirschen2018fundamentals}. For instance, generators participating in ISO New England's market offer a stepwise cost function for reporting the marginal generation cost \citep{NewEngland}. The market is cleared with a linear cost objective. Other markets, such as PJM and MISO, ask generators to offer a piecewise linear function to reflect the marginal generation cost \citep{guo2019data}. Then, the market is cleared with a quadratic cost objective. The adoption of linear or quadratic cost functions is for ensuring the desirable market properties. Thus, in this work, we follow the current practice and use a quadratic cost function. In other settings such as the one considering the valve-point effect of the generators \citep{coelho2006combining,chiang2005improved}, the nonlinear cost function can be used.} Let $\bm{P}^\text{g}=[P_i^\text{g}]_{i \in \mathcal{N}}$, $\hat{\bm{l}}=[\hat{l}_i]_{i \in \mathcal{N}}$ denote the vector formed by the nodal generation power and the nodal demands. The formulation of ED problem, which does not include the temporal coupling constraints and can be solved at each single-period independently, is as follows: 
\begin{subequations}\label{1}
\begin{align} \mathop{\min}_{\bm{P}^\text{g}} \quad & \frac{1}{2}\bm{P}^{\text{g}\top}\bm{Q}\bm{P}^{\text{g}}+\bm{q}^\top\bm{P}^{\text{g}}+\sum_{i\in\mathcal{N}}c_i^\text{g}\label{1(a)}\\ 
   s.t. \quad &\sum_{i=1}^{|\mathcal{N}|}P_i^\text{g}=\sum_{i=1}^{|\mathcal{N}|}\hat{l}_i:\gamma\label{1(b)}
    \\ 
    &
    -\bm{f}\leq \bm{H}(\bm{P}^\text{g}-\hat{\bm{l}})\leq\bm{f}:\bm{\mu}_1,\bm{\mu}_2\label{1(c)}\\
    & \bm{P}^{\text{min}} \leq \bm{P}^\text{g} \leq \bm{P}^{\text{max}}\label{1(d)}: 
    \bm{\psi}_1,\bm{\psi}_2;
\end{align}
\end{subequations}
where 
\begin{align*}
\bm{Q} =
  \begin{bmatrix}
    a_1^{\text{g}} & & \\
    & \ddots & \\
    & & a_{|\mathcal{N}|}^{\text{g}}
  \end{bmatrix},
\bm{q} =
  \begin{bmatrix}
    b_1^{\text{g}}\\
    \vdots\\ b_{|\mathcal{N}|}^{\text{g}}
  \end{bmatrix}
\end{align*}
Eq. \eqref{1(b)} is the power balance constraint. Given the shift factor matrix $\bm{H}\in\mathbb{R}^{|\mathcal{A}|\times|\mathcal{N}|}$ which maps the nodal power injections to the power flow on lines \citep{kundur2022power, 7478156}, the transmission constraint is presented in \eqref{1(c)} bounded by the vector of maximum transmission capacity $\bm{f}$. The generation capacity constraint regarding $\bm{P}^\text{g}$ is given in \eqref{1(d)}, where $\bm{P}^{\text{min}}=[P^{\text{min}}_i]_{i\in \mathcal{N}},\bm{P}^{\text{max}}=[P^{\text{max}}_i]_{i\in \mathcal{N}}$ are the vectors regarding the minimum and maximum output power of generators. The dual variables are given behind the colons. 

% \textcolor{black}{
% \begin{subequations}\label{dualproblem}
% \begin{align} &\gamma^*,\bm{\mu}_1^*,\bm{\mu}_2^*,\bm{\psi}_1^*,\bm{\psi}_2^*=\mathop{\arg\max}_{\gamma,\bm{\mu}_1,\bm{\mu}_2,\bm{\psi}_1,\bm{\psi}_2}-\frac{1}{2}(\bm{q}^\top-\gamma\bm{1}_{|\mathcal{N}|}^\top-\bm{\mu}_1^\top\bm{H}+\bm{\mu}_2^\top\bm{H}-\bm{\psi}_1^\top+\bm{\psi}_2^\top)\bm{Q}^{-1}(\bm{q}-\nonumber\\
% &\quad \gamma\bm{1}_{|\mathcal{N}|}-\bm{\mu}_1\bm{H}+\bm{\mu}_2\bm{H}-\bm{\psi}_1+\bm{\psi}_2)+\sum_{i\in\mathcal{N}}c_i^\text{g}+(\gamma\bm{1}_{|\mathcal{N}|}^\top+\bm{\mu}_1^\top\bm{H}-\bm{\mu}_2^\top\bm{H})\bm{l}\nonumber\\
% &\quad-\bm{\mu}_1^\top\bm{f}-\bm{\mu}_2^\top\bm{f}+\bm{\psi}_1^\top\bm{P}^{\text{min}}-\bm{\psi}_2^\top\bm{P}^{\text{max}}\\ 
%     &s.t. \bm{\mu}_1 \geq 0 ,\bm{\mu}_2 \geq 0,\bm{\psi}_1 \geq 0,\bm{\psi}_2 \geq 0
% \end{align}
% \end{subequations}}

 Let $\lambda_i$ be the LMP at bus $i$. By computing the gradient of the  Lagrangian associated with \eqref{1} with respect to the demand vector $\hat{\bm{l}}$, the vector of LMP $\bm{\lambda}=[\lambda_i]_{i \in \mathcal{N}}$  can be expressed as~\citep{kirschen2018fundamentals},
\begin{equation}\label{2}
  \bm{\lambda}=\gamma^*\cdot\bm{1}_{|\mathcal{N}|} - \bm{H}^\top(\bm{\mu}_2^*-\bm{\mu} _1^*),
\end{equation}
where $\bm{1}_{|\mathcal{N}|}$ is an all-one column vector with dimension of $|\mathcal{N}|$, and $\gamma^*,\bm{\mu}_2^*,\bm{\mu}_1^*$ are the values of the optimal dual solutions. In the literature, it has been shown that the LMP can be decomposed as two components, namely the marginal energy component and the marginal congestion component \citep{radovanovic2019holistic}. Concretely, if there is no congestion (which means the line flow is within the capacity $\bm{f}$), the optimal dual solutions $\bm{\mu}_2^*,\bm{\mu}_1^*$ are all zero. Therefore, the nodal LMPs all equal to $\gamma^*$, which is the marginal power component reflecting the marginal cost of power at the reference bus. If some lines are congested, which means some inequalities in \eqref{1(c)} are active, then $\bm{\mu}_2^*-\bm{\mu}_1^* \neq \bm{0}$. An extra term $-\bm{H}^\top(\bm{\mu}_2^*-\bm{\mu} _1^*)$ is added and named as the marginal congestion component, which makes the nodal LMPs become different. In this line, a price spike may happen because of the added marginal congestion component. We note that since the power losses are not included in the ED problem \eqref{1}, the formulation of LMP in \eqref{2} does not have the corresponding components. Such a formulation is applicable to a transmission network \citep{ji2016probabilistic,7478156}.

\section{Methodological Components}
In this section, to formulate the spike prices mitigation problem under the DR framework with selected nodes, we first introduce the underlying physical mechanism setup considered in the DR targeting problem in Section~\ref{sec:physical_DR}.  Then, we formulate the DR targeting problem in Section~\ref{sec:DR_targeting}.

\begin{figure}
  \centering
  \includegraphics[scale=0.45]{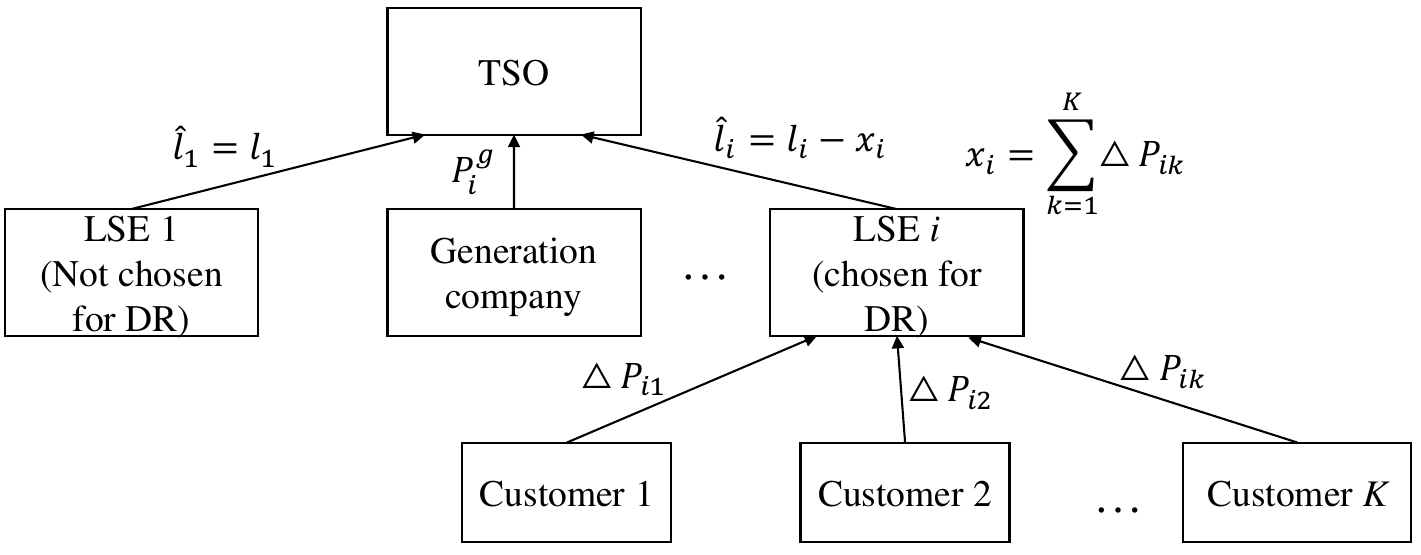}\\
  \caption{Illustration of the proposed DR targeting framework. When a DR event is triggered, TSO determines the a). targeted nodes  and  b). the corresponding demand reduction. The LSE connected to the targeted node (plotted in red) reduces the demand as required. }\label{Fig 1}
  \vspace{-1em}
\end{figure}

\subsection{The Physical Framework for DR Targeting}
\label{sec:physical_DR}
DR broadly refers to adapting power demand to achieve economic or grid-level goals. In this paper, we are interested in designing DR to improve market conditions, such as mitigating price spikes or reducing average electricity prices. The DR targeting framework is sketched in Fig. \ref{Fig 1}. In real-time energy market operation, TSO solves the ED problem based on the predicted nodal demands and derives LMP (\$/MWh) at each node. Following the practice in \citep{LEE2022103723}, a DR event is triggered once a price spike happens, which means the average LMP over all nodes is greater than a threshold (100 \$/MWh for instance in \citep{LEE2022103723}). To mitigate the price spike by reducing the average LMP, it is necessary to dispatch DR signals by reducing the demands at certain nodes.

\begin{figure}
  \centering
  \includegraphics[scale=0.5]{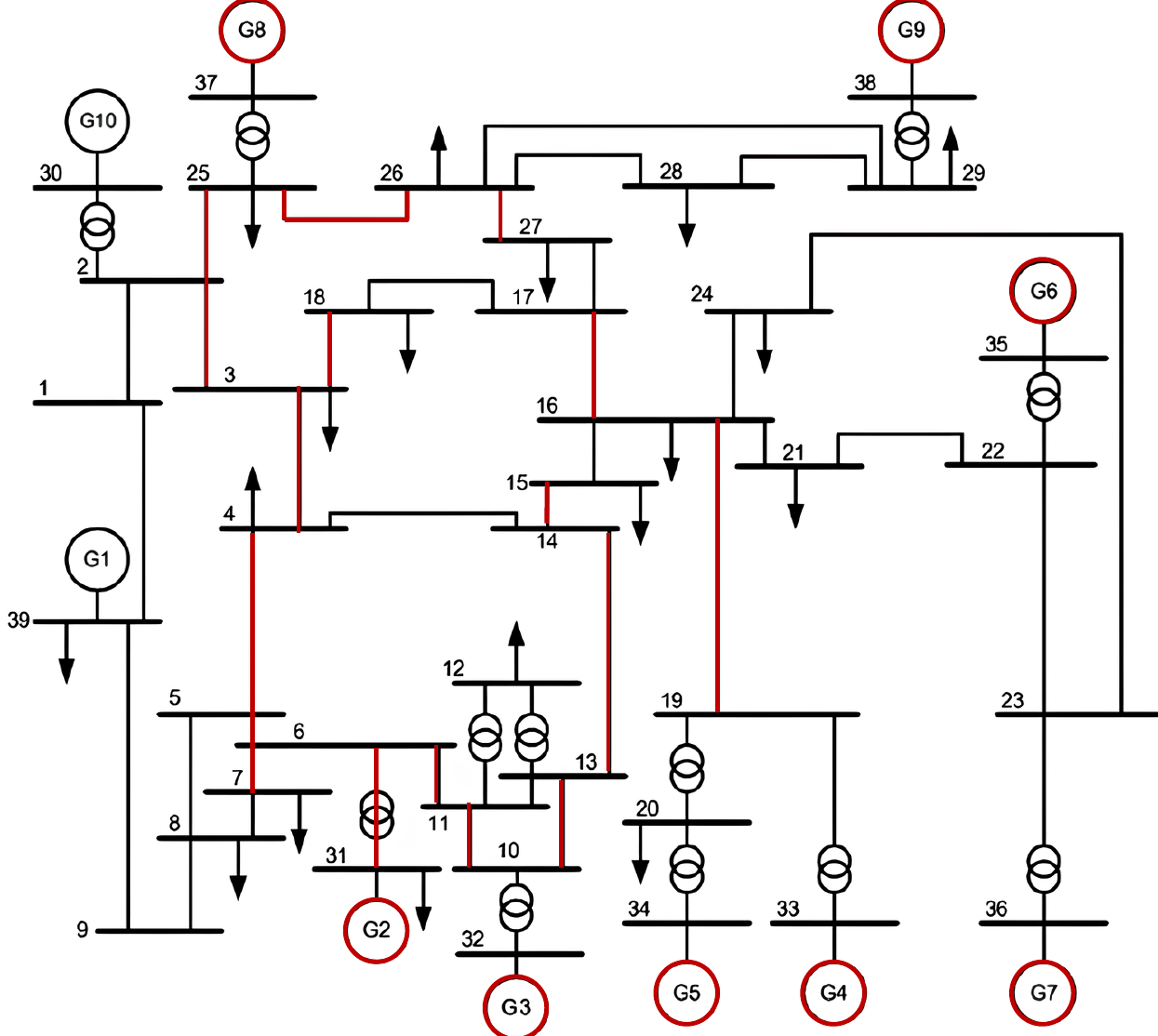}\\
  \caption{The system pattern before and after DR with improperly targeted node. Binding lines and nodes are shown in red.}\label{Fig 2}
\end{figure}

We now illustrate a concrete example to show that improper DR targeting can even result in an \emph{increase in the average LMP}. For example, Fig. \ref{Fig 2} shows the operation pattern before and after DR of the New England IEEE 39-bus system, where the red lines and circles display the binding lines and generators, if the node of the DR location is improperly chosen. Concretely, now we assume the operator asks to reduce the demand of 44.8 \unit{MW} only at node 25. Although the system pattern is not changed after DR, the inappropriate node selection results in the average LMP increasing from 105.6303 \unit{\$/MWh} to 106.1503 \$/MWh.  This is contradictory to the operating objectives of DR program. Therefore, properly choosing locations for implementing DR is necessary, so as to avoid such adverse effects. And in this work, we assume that TSO shoulders the responsibility of identifying the appropriate set of nodes to implement DR \citep{dehnavi2016determining}. They also need to determine the corresponding nodal demand reduction. The LSE, which is connected to the DR-targeted node, collects the DR resources from its served customers to fulfill the targeted demand reduction. Meanwhile, the  load level remains unchanged for those LSEs nodes not chosen for DR.

\subsection{The Formulation of DR Targeting}
\label{sec:DR_targeting}
In this subsection, we formulate the DR targeting problem for mitigating the price spike, which changes the nodal demands by DR, for the seek of reducing the average LMP. Let the binary vector $\bm{v}=[v_i]_{i \in \mathcal{N}}$ denote DR/non-DR locations, where $v_i=1$ means the node $i$ is chosen as the DR location, and $v_i=0$ means the node $i$ is not chosen. Take the example of IEEE 39-bus system in Fig. \ref{Fig 2}, only node 25 is chosen to implement DR, then $v_{25}=1$, and $v_i=0, \; \forall i \in \mathcal{N}/\{25\}$. Let the continuous vector $\bm{x}=[x_i]_{i \in \mathcal{N}}$ denote the nodal demand reduction whose upper bound is $\bar{\bm{x}}=[\bar{x}_i]_{i\in \mathcal{N}}$ which gives the maximum allowable demand reduction at each node,
% $\bm{\tau}=[w_i]_{i \in \mathcal{N}}$  is the weight vector, whose physical meaning can be the corresponding load reduction cost,
and $\hat{\bm{l}}$ denote the nodal load vector after DR. We formulate the DR targeting problem, which is given as
\begin{subequations}\label{3}
\begin{align} \mathop{\min}_{\bm{v},\bm{x}} \quad & \bm{\tau}^\top\bm{x}\label{3(a)}\\ 
    s.t. \quad &\sum_{i \in \mathcal{N}}v_i \leq K \label{3(b)}
    \\ 
    & \ 
    \bm{0} \leq \bm{x} \leq \bar{\bm{x}}\cdot \bm{v}\label{3(c)}\\
    & \bm{\hat{l}}=\bm{l}-\bm{x}\label{3(d)}\\
    & \bm{\lambda}=\gamma^*\cdot\bm{1}_{|\mathcal{N}|} - \bm{H}^\top(\bm{\mu}_2^*-\bm{\mu} _1^*)\label{lmp}\\
    &-\epsilon \leq \frac{1}{|\mathcal{N}|}\sum_{i \in \mathcal{N}}\lambda_i-\hat{\lambda}\leq \epsilon \label{3f}\\
    & \gamma^*,\bm{\mu}_2^*,\bm{\mu} _1^*\in \mathop{\arg\max}_{\gamma,\bm{\mu}_1\geq 0,\bm{\mu}_2\geq 0,\bm{\psi}_1\geq 0,\bm{\psi}_2\geq 0}-\frac{1}{2}(\bm{q}^\top-\gamma\bm{1}_{|\mathcal{N}|}^\top-\bm{\mu}_1^\top\bm{H}+\bm{\mu}_2^\top\bm{H}-\bm{\psi}_1^\top+\bm{\psi}_2^\top)\bm{Q}^{-1}\nonumber\\&(\bm{q}- \gamma\bm{1}_{|\mathcal{N}|}-\bm{\mu}_1\bm{H}+\bm{\mu}_2\bm{H}-\bm{\psi}_1+\bm{\psi}_2)+\sum_{i\in\mathcal{N}}c_i^\text{g}+(\gamma\bm{1}_{|\mathcal{N}|}^\top+\bm{\mu}_1^\top\bm{H}-\bm{\mu}_2^\top\bm{H})\bm{\hat{l}}\nonumber\\
&-\bm{\mu}_1^\top\bm{f}-\bm{\mu}_2^\top\bm{f}+\bm{\psi}_1^\top\bm{P}^{\text{min}}-\bm{\psi}_2^\top\bm{P}^{\text{max}} 
\label{3(e)}
\end{align}
\end{subequations}
where the objective function \eqref{3(a)} minimizes the linear demand reduction cost \citep{ming2018revenue,zhong2012coupon}, with $\bm{\tau}$ as the cost for per unit demand reduction. Other demand reduction costs such as the quadratic functions are also applicable and solvable using our proposed method.
% We normalize the LMPs and demand reduction into unitless values, by dividing them with $\lambda_{base}$ and $x_{base}$, respectively. Also, since the LMPs after DR will be smaller than that before DR, the normalized LMP will be in the range of $[0,1]$. So does the normalized demand reduction.
% The objective function \eqref{3(a)} is composed of minimizing the difference between the normalized average LMP and the normalized reference value $\frac{\hat{\lambda}}{\lambda_{base}}$, along with minimizing the normalized total demand reduction. 
% The first objective term is designed out of the two reasons. Firstly, the average LMP, which is the results of the energy dispatch problem in \eqref{1}, can be reduced to a preset value $\hat{\lambda}$, which mitigates the price spike. In  the case where $\hat{\lambda}$ is set to zero, the goal is to  minimize the average LMP as much as possible. Secondly, the quadratic term makes the LMPs across the network close to each other, which ensures no congestion happens after DR by Proposition \ref{prop1}.
Constraint \eqref{3(b)} limits that the number of the targeted nodes should be less than or equal to a preset integer $K$. Such constraint is added since in practice, only limited number of nodes are eligible to implement DR. Constraint \eqref{3(c)} bounds the demand reduction of the targeted node between 0 and $\bar{x}_i$. \eqref{3(d)} gives the new load vector $\bm{\hat{l}}$, which is the original load vector $\bm{l}$ minus the demand reduction $\bm{x}$. \eqref{lmp} gives the LMPs formed by the optimal dual solutions of ED, and \eqref{3f} limits the deviation of average LMP from the reference value $\hat{\lambda}$ to be within $\epsilon$. Given the new load vector $\bm{\hat{l}}$ as an input parameter, \eqref{3(e)} is the dual problem of \eqref{1}, which returns the optimal dual solutions that form LMP $\bm{\lambda}$. The $\mathop{\arg\max}$ operator makes \eqref{3(e)} the lower-level problem in \eqref{3}. The upper level is formed by \eqref{3(a)}-\eqref{3f}.

The dual problem of ED in \eqref{3(e)} maximizes the social welfare regarding the generation companies (the producers) and the LSEs (the consumers). The DR targeting problem seeks the optimal nodal demands $\bm{\hat{l}}$, to lower the price. The relationship between the nodal demands $\bm{\hat{l}}$ and the LMPs $\bm{\lambda}$, which is key to this problem, is implicitly given in the KKT conditions of \eqref{3(e)}. The nonconvex KKT conditions result in the nonconvex relationship between the nodal demands $\bm{\hat{l}}$ and the LMPs $\bm{\lambda}$, and therefore the nonconvex relationship between the demand reduction $\bm{x}$ and the LMPs $\bm{\lambda}$. Therefore, the constraint \eqref{3f} is a nonconvex one with respect to the demand reduction $\bm{x}$. This makes \eqref{3} a mixed integer program with the nonconvex constraint, which is very difficult to solve.

Fortunately, the relationship between the optimal dual solutions and the parameters of a QP program can be explicitly derived, and is shown to be  bijective and piecewise linear \citep{Grancharova2012}. Essentially, as the LMP vector $\bm{\lambda}$ is a specific linear transformation of the optimal dual solutions, the relationship between the parameters in \eqref{1}, such as the vector of load $\bm{\hat{l}}$, and the LMP vector $\bm{\lambda}$ is also a bijection and a piecewise linear function. And we note that such a piecewise linear function is not necessarily a convex function. In this work, we propose to describe such a relationship explicitly for replacing the lower-level program \eqref{3(e)}, and denote it as $\bm{\lambda}=\pi(\bm{\hat{l}})$. We will solve this problem individually for each segment and obtain the solutions yielding the minimum optimal objective. In this way, each linear segment of the function $\pi(\bm{\hat{l}})$ will be substituted into \eqref{3f}, leading to a mixed-integer linear program. Such a problem can be efficiently solved using off-the-shelf solvers. In the following sections, we will detail the solution process for $\pi(\cdot)$ and how it helps find DR targets.

 \begin{remark}
  We note that without considering the binary variable $\bm{v}$ and the corresponding constraint in \eqref{3(b)}, the problem in \eqref{3} reduces to the formulation for universal DR, where all nodes reduce the demands for mitigating the price spike.   
 \end{remark}

\begin{remark}
\textcolor{black}{We note that the iterative solution approach for the bilevel program, as adopted in \citep{wang2018equilibrium,lv2020power}, is difficult to apply to problem \eqref{3}. Specifically, this approach solves the upper-level and lower-level problems iteratively and separately. The locational marginal prices (LMPs) are determined by the lower level and are thus fixed parameters at the upper level, making them uncontrollable. Consequently, the constraint regarding the average LMP at the upper level may be violated, and price spikes cannot be effectively mitigated.}
 \end{remark}

\section{Price-demand Relationship Derivation via  Multi-parametric Quadratic Programming}

In this section, we derive the relationship between the vector of LMP and the load vector via multi-parametric quadratic programming theory. An illustrative example is presented in Section~\ref{sec:toy}, to intuitively present the policy in one-dimensional space. And the policy derivation algorithm for ED problem is given in Section~\ref{sec:policy}.

\subsection{An Illustrative Example}
\label{sec:toy}
A one-bus, two-generator example is presented to motivate finding the relationship between the LMP and the load numerically, and such analysis can be generalized to any large-scale networks. Consider a simple one-bus system with two generators serving the load $\hat{l}$. The generation cost $C_\text{g}(P_1^\text{g})=\frac{1}{2}a_1^\text{g} \cdot (P_1^\text{g})^2+c_1^\text{g}$ of G1 is cheaper than that of G2, i.e., $C_\text{g}(P_2^\text{g})=\frac{1}{2}a_2^\text{g} \cdot (P_2^\text{g})^2+c_2^\text{g}$, under the same generation level. The ED problem becomes
\begin{subequations}\label{5}
\begin{align} \mathop{\min}_{P_i^\text{g}} \quad &  \sum_{i=1}^{2}C_\text{g}(P_i^\text{g})\label{5(a)}\\ 
    s.t. \quad &\sum_{i=1}^{2}P_i^\text{g}=\hat{l}\label{5(b)}
    \\ 
    & 0 \leq P_i^\text{g} \leq P^{\text{max}}_i,\forall i \in \{1,2\}\label{5(c)}.
\end{align}
\end{subequations}

\begin{figure}
  \centering
  \includegraphics[scale=0.6]{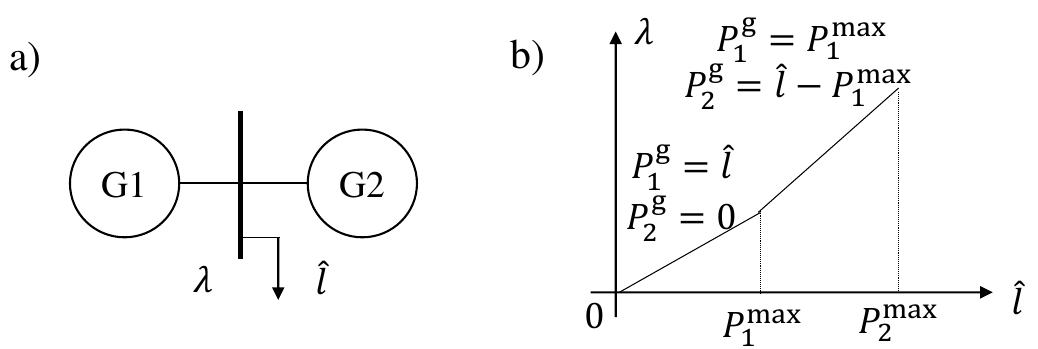}\\
  \caption{The policy curve of a single bus load with two generators.}\label{Fig 3}
\end{figure}

Fig. \ref{Fig 3} (b) plots LMP $\lambda$ versus load $\hat{l}$. The curve is piecewise linear, convex and increasing. Different segments of load corresponds to different affine functions with varying slopes as shown in Fig. \ref{Fig 3}. Specifically, as the generation cost of $P_1^\text{g}$ is cheaper, the slope of the affine line in the first segment is related with $\frac{\partial C_\text{g}(P_1^\text{g})}{\partial P_1^\text{g}}$, i.e., $a_1^\text{g}$. When the power output of $P_1^\text{g}$ reaches the upper bound $P^{\text{max}}_1$, the generator $P_2^\text{g}$ begins to output power to serve the load. Therefore, the slope of the affine line in the second segment is  $a_2^\text{g}$. The LMP is continuous within each line segment, and discontinuous between the segments.
The example in the one-bus system illustrates the LMP can be described by a piecewise linear function over the scalar load variable $\hat{l}$. Yet the price-demand relationship of a system with multiple nodes along with the line constraints is more complex, as the region of loads is systematically characterized by high-dimensional polytopes. In the next subsection, we will show how to find such a relationship via multi-parametric quadratic programming theory.

\subsection{The Derivation of Price-demand Relationship}
\label{sec:policy}
Given the ED problem in \eqref{1}, the goal of this subsection is to derive the function between the LMP vector $\bm{\lambda}$ and the load vector $\hat{\bm{l}}$. Let $\bm{I} \in \mathbb{R}^{|\mathcal{N}| \times |\mathcal{N}|}$ be the identity matrix. Before going to the details, we firstly rewrite the ED problem in \eqref{1} into the compact form of a QP problem
\begin{subequations}\label{6}
\begin{align} \mathop{\min}_{\bm{P^g}} \quad & \frac{1}{2}(\bm{P}^\text{g})^{\top}\bm{Q}(\bm{P}^\text{g})+\bm{q}^{\top}\bm{P}^\text{g}+\sum_{i\in\mathcal{N}}c_i^\text{g}
\label{6(a)}\\ 
    s.t. \quad & \bm{1}_{|\mathcal{N}|}^{\top}\bm{P}^\text{g}=\bm{1}_{|\mathcal{N}|}^{\top}\hat{\bm{l}}:\gamma\label{6(b)}
    \\ 
    \quad \ 
    &\bm{A}\bm{P}^\text{g}\leq\bm{b}:\bm{\mu}\label{6(c)}\\
    \quad \ 
    &\bm{S}\bm{P}^\text{g}\leq\bm{h}:\bm{\psi}\label{6(d)},
\end{align}
\end{subequations}
where 
\begin{align*}
\bm{A}=
\begin{bmatrix}
\bm{H}\\
-\bm{H}\\
\end{bmatrix},
\bm{b}=
\begin{bmatrix}
\ \bm{f}+\bm{H}\bm{l}\ \\
\ \bm{f}-\bm{H}\bm{l}
\end{bmatrix},
\bm{S} =
  \begin{bmatrix}
  \bm{I}\\
  -\bm{I}\\
  \end{bmatrix},
\end{align*}

\begin{align*}
  \bm{\mu} =
  \begin{bmatrix}
  \bm{\mu}_2\\
  \bm{\mu}_1\\
  \end{bmatrix},
\bm{\psi} =
  \begin{bmatrix}
  \bm{\psi}_2\\
  \bm{\psi}_1\\
  \end{bmatrix}, 
  \bm{h}_1 =
  \begin{bmatrix}
  -P^{\text{min}}_1\\
  \vdots\\
  -P^{\text{min}}_{|\mathcal{N}|}\\
  \end{bmatrix},
  \bm{h}_2 =
  \begin{bmatrix}
  P^{\text{max}}_1\\
  \vdots\\
  P^{\text{max}}_{|\mathcal{N}|}\\
  \end{bmatrix},
    \bm{h} =
  \begin{bmatrix}
  \bm{h}_2\\
  \bm{h}_1\\
  \end{bmatrix}.
\end{align*}

We now proceed with the following theorem to characterize the relationship between input parameters and the optimal solutions of a QP.

\textbf{Theorem 1 \citep{Grancharova2012}:} Consider the parameters of a QP problem belong to a convex set and the parameters $\bm{Q}$ is positive definite. The optimal solution of a QP problem is  given by a conditional piecewise linear function of the varying parameters in the constraints.

\textbf{Theorem 1} shows that given an initial convex region $\mathcal{L}$ of all feasible vectors of loads, the relationship between the optimal primal and dual solutions of \eqref{6} and load is piecewise linear. Therefore, we can have the corollary from the \textbf{Theorem 1} that in a subregion of load, the mapping between the optimal solution and the load is an affine function. Therefore, let $M$ denote the number of subregions characterized the initial region $\mathcal{L}$. For the $m_{th},\forall m=1,...,M$ subregion in $\mathcal{L}$, we can have the following proposition for the QP problem in \eqref{6}.

\begin{proposition}\label{prop2}
 Consider the QP problem \eqref{6}. Let $\bm{P}^\text{g}_m(\hat{\bm{l}}),\gamma_m(\hat{\bm{l}}),\bm{\mu}_m(\hat{\bm{l}}),\bm{\psi}_m(\hat{\bm{l}})$ denote the mapping between the load $\hat{\bm{l}}$ and the optimal solutions of primal and dual variables, $\Tilde{\bm{P}^\text{g}},\Tilde{\gamma},\Tilde{\bm{\mu}},\Tilde{\bm{\psi}}$ denote the optimal primal and dual solutions under a given load vector $\Tilde{\bm{l}}$. The calculation of $\bm{P}^\text{g}_m(\hat{\bm{l}}),\gamma_m(\hat{\bm{l}}),\bm{\mu}_m(\hat{\bm{l}}),\bm{\psi}_m(\hat{\bm{l}})$, in the $m_{th}$ subregion that $\hat{\bm{l}}=\Tilde{\bm{l}}$ resides, is
\begin{equation}\label{7}
  \begin{bmatrix}
  \bm{P}^\text{g}_m(\hat{\bm{l}})\\
  \gamma_m(\hat{\bm{l}})\\
  \bm{\mu}_m(\hat{\bm{l}})\\
  \bm{\psi}_m(\hat{\bm{l}})
  \end{bmatrix} =
  \begin{bmatrix}
  \Tilde{\bm{P}^\text{g}}\\
  \Tilde{\gamma}\\\Tilde{\bm{\mu}}\\\Tilde{\bm{\psi}}
  \end{bmatrix}+(\bm{M}_0)^{-1}\cdot \bm{N}_0\cdot (\hat{\bm{l}}-\Tilde{\bm{l}}),
\end{equation}
where $\bm{M}_0 \in \mathbb{R}^{(3|\mathcal{N}|+1+2|\mathcal{A}|)\times(3|\mathcal{N}|+1+2|\mathcal{A}|)}$ and $\bm{N}_0 \in \mathbb{R}^{(3|\mathcal{N}|+1+2|\mathcal{A}|)\times|\mathcal{N}|}:$
\begin{align*}
&\bm{M}_0 =
  \begin{bmatrix}
  \bm{Q} & \bm{1}_{|\mathcal{N}|} & \bm{A}^\top & \bm{S}^\top\\
\bm{1}_{|\mathcal{N}|}^\top & 0 & 0 &0\\
D(\Tilde{\bm{\mu}})\bm{A} & \bm{0} &D(\bm{A}\Tilde{\bm{P}^\text{g}}-\bm{b}) & \bm{0}\\
D(\Tilde{\bm{\psi}})\bm{S} & \bm{0} & \bm{0} & D(\bm{S}\Tilde{\bm{P}^\text{g}}-\bm{h})\\
  \end{bmatrix},
\bm{N}_0= \begin{bmatrix}
\bm{0}\\
\bm{1}_{|\mathcal{N}|}^\top\\
D(\Tilde{\bm{\mu}})
\begin{bmatrix}
\bm{H}\\
-\bm{H}
\end{bmatrix}\bm{I}\\
\bm{0}
\end{bmatrix}.
\end{align*}
$D(\cdot)$ creates a diagonal matrix from a vector. 
\end{proposition}
\begin{proof}
    See in the \ref{Appendix A}.
\end{proof}
% The proof of Proposition \ref{prop2} is based on the KKT conditions of \eqref{6} and given in \ref{Appendix A}.
Once \eqref{7} is obtained, we can find the function of the optimal primal and dual solutions with respect to $\hat{\bm{l}}$ in this subregion. Based on \eqref{2}, the local policy $\pi_m(\hat{\bm{l}})$ in this subregion can be derived as
\begin{equation}\label{8}
   \pi_m(\hat{\bm{l}})=\gamma_m(\hat{\bm{l}})\cdot\bm{1}_{|\mathcal{N}|} - \bm{H}^\top[\bm{\mu}_{2,m}(\hat{\bm{l}})-\bm{\mu}_{1,m}(\hat{\bm{l}})].
\end{equation}
where 
\begin{align*}
\bm{\mu}_m(\hat{\bm{l}})=
\begin{bmatrix}
  \bm{\mu}_{2,m}(\hat{\bm{l}})\\
  \bm{\mu}_{1,m}(\hat{\bm{l}})\\
  \end{bmatrix}
\end{align*}

Recall in the DR targeting task, we are interested in finding LMPs given varying demand vectors. Such a task then motivates us to firstly identify $\pi_m(\hat{\bm{l}})$  for a specific $\Tilde{\bm{l}}$ using \eqref{8}, and then to work through all subregions that have distinct policy parameters defined by \eqref{7}. The subregion of $\Tilde{\bm{l}}$ where \eqref{7} holds is defined as the critical region $\mathcal{B}_m(\Tilde{\bm{l}})$. Since \eqref{7} is derived by the KKT conditions, the optimality and feasibility conditions given in the KKT conditions remain invariant within the critical region $\mathcal{B}_m(\Tilde{\bm{l}})$. In other words, the active inequalities and the inactive inequalities remain invariant. Let $\Tilde{\bm{A}},\Tilde{\bm{b}}$, $\Tilde{\bm{S}}$, and $\Tilde{\bm{h}}$ denote the coefficients corresponding to the inactive inequalities, and $\Tilde{\bm{\mu}}_m(\hat{\bm{l}}),\Tilde{\bm{\psi}}_m(\hat{\bm{l}})$
be part of the $\bm{\mu}_m(\hat{\bm{l}}),\bm{\psi}_m(\hat{\bm{l}})$ corresponding to the active inequalities. Feasibility is ensured by substituting $\bm{P}^\text{g}_m(\hat{\bm{l}})$ into the inactive inequalities and the equality in \eqref{6(b)}, whereas the optimality condition is given by $\Tilde{\bm{\mu}}_m(\hat{\bm{l}}) \geq 0,\Tilde{\bm{\psi}}_m(\hat{\bm{l}}) \geq 0$. The critical region $\mathcal{B}_m(\Tilde{\bm{l}})$ is defined as
\begin{equation}\label{9}
\begin{split}
\Tilde{\mathcal{B}}_m(\Tilde{\bm{l}})= & \{\Tilde{\bm{A}}\bm{P}^\text{g}_m(\hat{\bm{l}}) \leq \Tilde{\bm{b}},\Tilde{\bm{S}}\bm{P}^\text{g}_m(\hat{\bm{l}}) \leq \Tilde{\bm{h}},\bm{1}_{|\mathcal{N}|}^{\top}\bm{P}^\text{g}_m(\hat{\bm{l}})=\bm{1}_{|\mathcal{N}|}^{\top}\hat{\bm{l}}, \Tilde{\bm{\mu}}_m(\hat{\bm{l}}) \geq 0,\Tilde{\bm{\psi}}_m(\hat{\bm{l}}) \geq 0,\mathcal{L}\};\\
 \mathcal{B}_m(\Tilde{\bm{l}})=&\triangle \{\Tilde{\mathcal{B}}_m(\Tilde{\bm{l}})\},
\end{split}
\end{equation}
where $\triangle \{\cdot\}$ is defined as an operator which removes redundant constraints; see \citep{Gal+2010}. Once $\mathcal{B}_m(\Tilde{\bm{l}})$ has been obtained, the rest of region $\mathcal{N}(\Tilde{\bm{l}})$ is obtained; see \citep{Dua:2000}:

\begin{equation}\label{10}
    \mathcal{N}(\Tilde{\bm{l}})=\mathcal{L}-\mathcal{B}_m(\Tilde{\bm{l}}).
\end{equation}
Given the remaining regions $\mathcal{N}(\Tilde{\bm{l}})$, the local policy along with the corresponding critical region can then be obtained with the similar way in Proposition \ref{prop2} iteratively. The algorithm terminates when there are no more regions to explore. The main steps of the algorithm are summarized in Algorithm \ref{alg1}. 

To sum up, the policy $\pi(\hat{\bm{l}})=\{\pi_m(\hat{\bm{l}})\}_{m=1}^M$ on the initial convex region $\mathcal{L}$ of feasible load vector is a piecewise linear function, where $M$ is the number of critical regions, and the local function $\pi_m(\hat{\bm{l}})$ is affine on each critical region which can be characterized by a high-dimensional polytope. With this representation of the policy $\pi(\hat{\bm{l}})$, the DR targeting problem formulated in \eqref{3} can be solved separately on each critical region with the policy represented by an affine function. This gives a mixed integer program with linear constraints, which can be solved very efficiently. A constraint regarding the space of the critical region is also added to the original problem. We will illustrate the overall solution strategy in the next Section.

\begin{algorithm}[h]
	%\textsl{}\setstretch{1.8}
	%\renewcommand{\algorithmicrequire}{\textbf{Input:}}
	\caption{The derivation of policy $\pi(\hat{\bm{l}})$ via critical region identification}
	\label{alg1}
	\begin{algorithmic}[1]
    \Require{The initial convex region of all feasible vector of loads $\mathcal{L}$}
    \State{In the given region of $\mathcal{L}$, solve \eqref{6} by treating $\hat{\bm{l}}$ as a free variable to obtain a feasible point $\Tilde{\bm{l}}$.}
    \State{Fix $\hat{\bm{l}}=\Tilde{\bm{l}}$ and solve \eqref{6} to obtain $\Tilde{\bm{P}^\text{g}},\Tilde{\gamma},\Tilde{\bm{\mu}},\Tilde{\bm{\psi}}$}.
    \State{Compute $(\bm{M}_0)^{-1}\cdot\bm{N}_0$ and obtain $\bm{P}^\text{g}_m(\hat{\bm{l}}),\gamma_m(\hat{\bm{l}}),\bm{\mu}_m(\hat{\bm{l}}),\bm{\psi}_m(\hat{\bm{l}})$ from \eqref{7}.}
    \State{Compute the policy $\pi_m(\hat{\bm{l}})$ from \eqref{8}.}

    \State{Obtain the critical region $\mathcal{B}_m(\Tilde{\bm{l}})$ as defined in \eqref{9}.}
    \State{Obtain $\mathcal{N}(\Tilde{\bm{l}})$ using  \eqref{10}.}
    \State{If no more regions to explore, go to next step, otherwise given $\mathcal{N}(\Tilde{\bm{l}})$, solve \eqref{6} by treating $\hat{\bm{l}}$ as a free variable to obtain a new feasible point $\Tilde{\bm{l}}$, and then go to Step 2.}
    \State{Collect all the solutions and unify the critical regions having the same solution  to obtain a compact representation.}
	\end{algorithmic}

\end{algorithm}

\section{Solution Strategy for DR Targeting Problem}

Essentially, given the derived price-demand relationship $\pi(\hat{\bm{l}})$, we find an explicit function between the electricity demand at any time-slot and the nodal LMP. We show a MILP formulation of DR targeting problem in this section. An acceleration strategy equipped with $\pi(\hat{\bm{l}})$ is introduced to efficiently solve such the DR targeting problem.

With the function $\pi(\hat{\bm{l}})$, we write each critical region in the form of polytope. For the $m_{th}$ critical region, the corresponding function $\pi_m(\hat{\bm{l}})$ is affine. Here, we denote the region of the polytope as $\bm{R}_m \cdot \hat{\bm{l}} \leq \bm{r}_m$. In this way, the DR targeting problem within this polytope is a MILP, given by
\begin{subequations}\label{11}
\begin{align}
\mathop{\min}_{\bm{v},\bm{x}}  \quad &\bm{\tau}^\top\bm{x}\\
s.t. \quad & \eqref{3(b)},\eqref{3(c)},\eqref{3(d)},\eqref{lmp},\eqref{3f}\nonumber\\
\quad \ & \bm{\lambda}=\pi_m(\bm{\hat{l}})\\
\quad \ &\bm{R}_m \cdot \bm{\hat{l}} \leq \bm{r}_m.\label{11(c)}
\end{align} 
\end{subequations}

Although we can solve the MILP problem for each polytope, and return the optimal solution of the problem whose optimal objective is the smallest among the $M$ problems, this practice is computationally expensive, as solving the MILP problem takes a relatively long time for multiple polytopes. Considering the fact that the allowable demand reduction is upper bounded, there are only a set of polytopes containing feasible solutions for the MILP problem. While these polytopes only take up a small portion among the $M$ problems, we can greatly reduce the time of solving the DR targeting by only working on \emph{feasible DR regions}. To accelerate the process, we propose to firstly judge the feasibility of \eqref{11} before solving it. As such, the time of solving the infeasible MILP problem is spared. Therefore, the relaxed version of \eqref{11} is firstly solved to help judge the feasibility, which is given by
\begin{subequations}\label{12}
\begin{align}
\mathop{\min}_{\bm{v},\bm{x}} \quad &\bm{\tau}^\top\bm{x}\\
s.t. \quad &\eqref{3(c)},\eqref{3(d)},\eqref{lmp},\eqref{3f}\nonumber\\
\quad \ & 0 \leq v_i \leq 1, \forall i \in \mathcal{N}\\
\quad \ & \bm{\lambda}=\pi_m(\bm{\hat{l}})\\
\quad \ &\bm{R}_m \cdot \bm{\hat{l}} \leq \bm{r}_m \label{12(d)},
\end{align} 
\end{subequations}
where the binary variable $v_i$ is relaxed to be a continuous variable in the range of $[0,1]$. As $v_i$ is a continuous variable in \eqref{12}, the cardinality constraint \eqref{3(b)} bounding the number of DR locations is omitted. If the relaxed problem in \eqref{12} is feasible, then solve the MILP problem in \eqref{11}; otherwise go to the problem on the next polytope. The pseudocode regarding the solution strategy is summarized in Algorithm 2.

\begin{algorithm}[h]
	%\textsl{}\setstretch{1.8}
	%\renewcommand{\algorithmicrequire}{\textbf{Input:}}
	\caption{The solution strategy for DR targeting}
	\label{alg2}
	\begin{algorithmic}[1]
    \Require{The derived policy $\pi(\hat{\bm{l}})$, which is defined on the region $\mathcal{L}$ formed by $M$ polytopes, and for the $m_{th}$ polytope, the policy is an affine function defined as $\pi_m(\hat{\bm{l}})$.}

    \For{$m =1,...,M$}
    \State{Given the policy $\pi_m(\hat{\bm{l}})$ and the corresponding polytope $\bm{R}_m \cdot \bm{\hat{l}} \leq \bm{r}_m$, solve the relaxed problem in \eqref{12}.}
    \State{If \eqref{12} is feasible, then solve the MILP problem in \eqref{11}. If the MILP problem in \eqref{11} is feasible, then store the optimal solution and objective in the buffer.}  
    \EndFor
    \State{Return the optimal solution whose optimal objective is the smallest in the buffer.}	\end{algorithmic}

\end{algorithm}

\section{Case Study}
\begin{figure}
  \centering
  \includegraphics[scale=0.7]{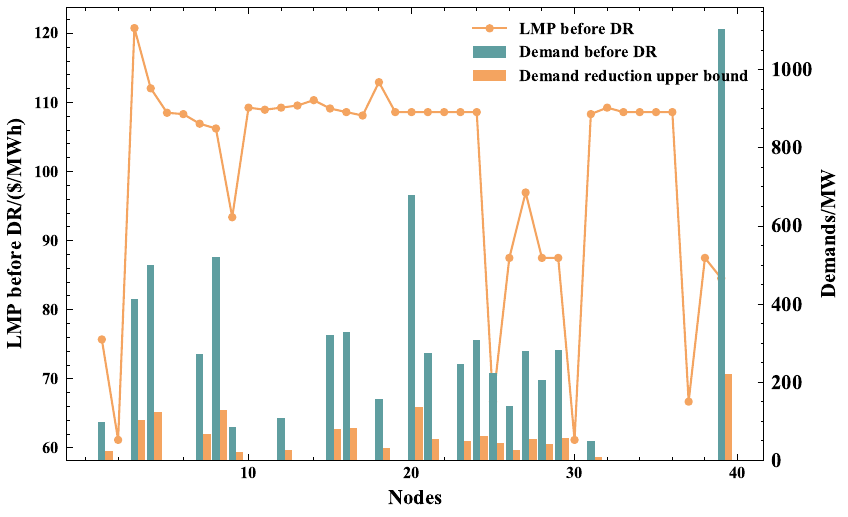}\\
  \caption{The nodal LMPs and nodal demand levels before DR, along with demand reduction upper bound.} \label{Fig 4}
\end{figure}

This section testifies the effectiveness of the proposed approach on New England IEEE 39-Bus system. We adapt the scale of parameters in the standard IEEE 39-Bus system. In addition, without the special statement, the nodal demand levels $\bm{l}$ and nodal LMPs before DR used in the subsequent subsections are the ones illustrated in Fig. \ref{Fig 4}, along with the demand reduction upper bound $\bar{\bm{x}}$. Concretely, $\bar{\bm{x}}$ is chosen as either 25\% or 20\% of the nodal demand levels $\bm{l}$. The demand reduction cost is set as 50 \$/MWh, for each entry in the vector $\bm{\tau}$. The value of reference price $\hat{\lambda}$ will be specified in the following discussions. Such a value is set in an iterative way. If the relaxed program in \eqref{12} is infeasible, we increase the value until the feasibility is reached. Here, we chose the approach implements DR at the nodes with the highest LMPs as the comparison approach. Concretely, an optimal power flow problem is formulated, for minimizing the generation cost and the demand reduction cost. In general, the comparison approach and ours diverge on two fronts. Firstly, the comparison approach solely determines demand reduction at specified locations, without rigorous optimization to identify DR locations. Secondly, it lacks explicit constraints to limit the reduction of average LMPs to the reference value. Consequently, the efficacy of leveraging demand response for price spike mitigation may not be realized efficiently. The detailed formulation is in \ref{Appendix B}.

The aim is to show the proposed approach (1) can reduce the average LMPs to the reference point compared to heuristics such as selecting DR locations with highest LMPs; (2) achieves efficient DR targeting under different parameter settings; (3) can achieve effective policies even when network parameters are inaccurate.
% Lastly, to showcase the proposed approach is also generalizable to other DR applications, we consider the case of load shifting and show the proposed approach can still effectively reduce the average LMPs.

 All simulations are implemented on the laptop with Intel®CoreTM i5-10210U 1.6 GHz CPU, and 8.00 GM RAM and based on the Multiparametric Toolbox \citep{MPT3}.

\subsection{Operational Advantage of the Proposed Approach}

In this subsection, we set the number of targeted DR nodes $K$ in \eqref{3(b)} as 5, and the acceptable average LMP deviation $\epsilon$ as 0.01 \$/MWh.
% Given that the LMP at each node can be viewed as an additional cost incurred when adding one more unit of nodal demand, it is a natural decision to consider reducing the demand at nodes with relatively high LMPs, for reducing the system average LMP.
The benchmark implements DR at the nodes with the highest LMP. Under such a node selection policy, \#3, \#4, \#12, \#15, \#18 are selected as locations for implementing DR. \textcolor{black}{We note that the selected DR locations also correspond to the locations where each unit of demand reduction causes the largest reduction in the average LMP.} 

The reference LMP, average LMP after DR, and total demand reduction cost are shown in Table \ref{Table 1}.
When the reference LMP is set as 68.75 \$/MWh, both approaches can lower the average LMP to match the reference value, with a deviation within $\epsilon$. Additionally, the proposed approach achieves a lower total demand reduction cost. When the reference LMP is set as 65 \$/MWh, the comparison candidate fails to decrease the average LMP to match the reference value, whereas the proposed approach succeeds in doing so. The inadequacy of the comparison candidate can be attributed to the improper selection of DR locations. The results reveal that the ideal nodes to implement DR are not necessarily coinciding with the nodes with the highest LMPs, which is reasonable as the comparison candidate produces a feasible solution to the proposed MILP problem in \eqref{11}, not an optimal one. The results demonstrate that the proposed approach can bring more advantages the operation, compared to the comparison candidate.

\begin{table}[h]
\caption{ 
The comparison between the proposed and the DR targeting method selecting the targeted nodes with the highest LMP.}\label{Table 1}
\begin{center}
\begin{tabular}{c | c c c}
\hline\hline 
 & \makecell[c]{Reference LMP\\ (\$/MWh)} &
    \makecell[c]{Average LMP\\ (\$/MWh)} & \makecell[c]{Total demand\\ reduction cost (\$)}\\
\hline
\makecell[c]{Proposed} & 68.75 &
    68.76 & 18152\\
\makecell[c]{Comparison} & 68.75 &
    68.75 & 18337\\
\hline
\makecell[c]{Proposed}    
 & 65 &
    65.01 & 30576\\
\makecell[c]{Comparison} & 65 &
    68.75 & 18337\\
\hline
\hline\hline
\end{tabular}
\end{center}
\end{table}

Furthermore, the targeted nodes of the two approaches and the corresponding demand reduction (shown in the parentheses) are displayed in Table \ref{Table 2}. The proposed approach only need to reduce demand at four nodes. Interestingly, except for the node \#4, the nodes with the highest LMPs are not selected as the DR locations by the proposed method. And the node with the relatively low LMP value is chosen, for example the node \#29. Also, the targeted nodes of the proposed method are not necessarily the load centers with heavy demand. For instance, the node \#29 is selected, whose demand is relatively small. The results further show that the intuitive wisdom is not necessarily the optimal one. And the rigorous examination is needed to target the proper nodes for more effective DR.

\begin{table}[h]
\caption{ The index of targeted nodes and corresponding demand reduction (in MW), when the reference LMP is set as 68.75 (\$/MWh) .}\label{Table 2}
\begin{center}
\begin{tabular}{c | c c c  c c}
\hline\hline  \makecell[c]{Proposed} & \#4(125) &
    \#16(82.25) & \#20(120) &
    \#29(35.78)\\
\hline
\makecell[c]{Comparison} & \#3(103) &
    \#4(125) & \#12(27.13) &
    \#15(80) &
    \#18(31.6) \\
\hline\hline
\end{tabular}
\end{center}
\end{table}

% \begin{figure}
%   \centering
%   % Requires \usepackage{graphicx}
%   \includegraphics[scale=0.6]{Fig5.pdf}\\
%   \caption{The comparison between the proposed and the DR targeting method selecting the targeted nodes with the highest LMP. }\label{Fig 5}
% \end{figure}

% \begin{figure}
%   \centering
%   % Requires \usepackage{graphicx}
%   \includegraphics[scale=0.7]{LMP_afterDRV2.pdf}\\
%   \caption{Nodal LMP of the proposed approach and comparison candidate after DR.  }\label{Fig 5}
% \end{figure}

\begin{figure}
  \centering
  \includegraphics[scale=0.6]{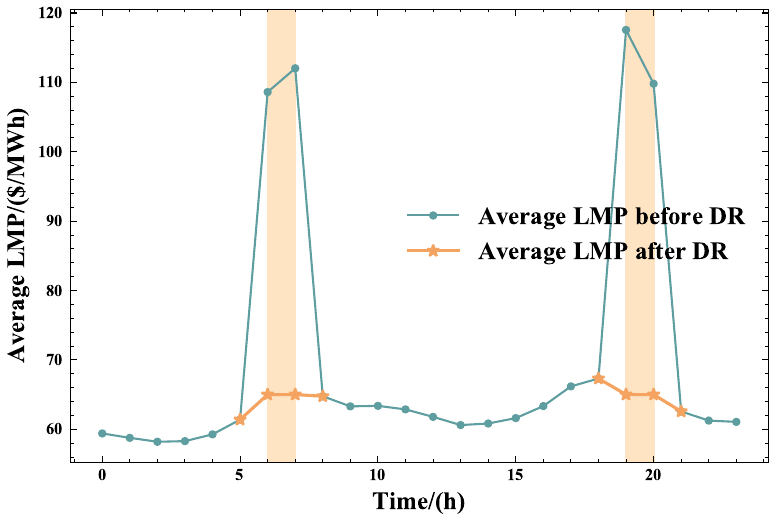}\\
  \caption{The average LMP before and after DR in a day. Shaded regions depict DR events with price spikes, where the total demand reduction at 7 am, 8 am, 7 pm, and 8 pm are 644, 662, 655, and 641 MW, respectively.}\label{Fig 6}
\end{figure}

To show the effectiveness of mitigating the price spike, the average LMP before and after DR in a day is shown in Fig. \ref{Fig 6}. The DR targeting is triggered once the average LMP is larger than 100 \$/MWh, as depicted in the shaded area in the figure. It is shown that  before implementing DR, the average LMP exhibits significant temporal fluctuations. After that, the profile of the average LMP flattens out and remains at a relatively low level.  

Also, by testing the proposed approach and the comparing candidate on different demand levels before DR, we aim to show that the proposed approach is superior to the method where DR is implemented at fixed nodes. Specifically, we test the approach on six kinds of different nodal demand levels before DR, which are denoted as $S_1-S_6$. The reference value of LMP is set as 65 \$/MWh. Those demand levels are generated by adding the Gaussian noise to the demand level shown in Fig. \ref{Fig 4}. The comparison candidate always selects the nodes \#3, \#4, \#12, \#15, \#18 as DR locations, whose corresponding LMPs are almost the highest in the system. The results regarding the reduced average LMP are given in Table \ref{Table 3}. 
 Among $S_1-S_6$, the proposed approach achieves a reduction of the average LMP to match the reference value, with a deviation less than $\epsilon$, whereas the comparison candidate fails to do so. For the proposed approach, the targeted nodes in $S_1-S_6$ are shown in Table \ref{Table 4}. The targeted nodes tend to vary with changes in demand levels prior to DR implementation. The results validate that the proposed approach can adapt the selection of DR targeted nodes to different demand levels, demonstrating its superiority over the approach of choosing fixed nodes. 

\begin{table}[h]
\caption{Average LMP (in \$/MWh) under different nodal demand levels before DR, when the reference LMP is set as 65 (\$/MWh) .}\label{Table 3}
\begin{center}
\begin{tabular}{c | c c c  c c c}
\hline\hline
 Demand Reduction   & $S_1$ & $S_2$ & $S_3$
    & $S_4$ & $S_5$& $S_6$\\
\hline    \makecell[c]{Proposed} & 65.01 &
    65.01 & 65.01 &
    65.01 &
    65.01 &
    65.01\\
\hline
\makecell[c]{Comparison} & 69.49 &
    69.79 & 69.54 &
    69.31 &
    69.25 & 69.02\\
\hline\hline
\end{tabular}
\end{center}
\end{table}

\begin{table}[h]
\caption{Targeted nodes of the proposed approach under different nodal demand levels before DR.}\label{Table 4}
\begin{center}
\begin{tabular}{c | c c c  c c}
\hline\hline
   $S_1$ & \#3 &
    \#4 & \#7 &
    \#8 &
    \#39\\
\hline
   $S_2$ & \#3 &
    \#4 & \#8 &
    \#20 &
    \#39\\
\hline
   $S_3$ & \#3 &
    \#8 & \#15 &
    \#20 &
    \#39\\
\hline
   $S_4$ & \#3 &
    \#4 & \#7 &
    \#8 &
    \#39\\
\hline
   $S_5$ & \#3 &
    \#4 & \#7 &
    \#8 &
    \#39\\
\hline
   $S_6$ & \#3 &
    \#4 & \#8 &
    \#29 &
    \#39\\
\hline\hline
\end{tabular}
\end{center}
\end{table}

\subsection{Investigation on Values of Different  Acceptable LMP Deviation}

\begin{figure}
  \centering
  \includegraphics[scale=0.7]{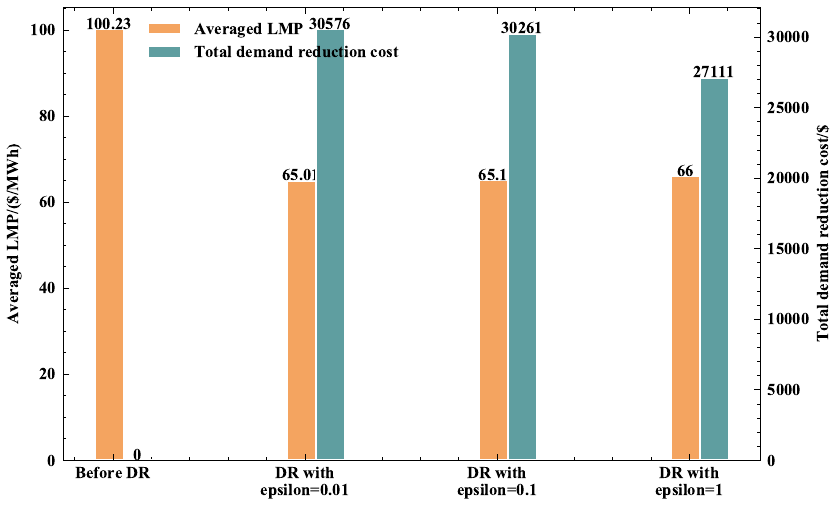}\\
  \caption{The nodal average LMP and total demand reduction cost under different values of acceptable LMP deviation.}\label{Fig 7}
\end{figure}

The value of the acceptable LMP deviation $\epsilon$ has an impact on the DR performance. Therefore, we investigate the influence of this parameter in this subsection. Specifically, the reference value of LMP is set as 65 \$/MWh, and the acceptable deviation $\epsilon$ is set as 0.01 \$/MWh, 0.1 \$/MWh, and 1 \$/MWh, respectively. The results in Fig. \ref{Fig 7} show that the reduced average LMP matches the reference value, with a deviation of $\epsilon$. With the increasing value of $\epsilon$, the average LMP increases, and the total demand reduction cost decreases. 
This outcome is reasonable, as less demand reduction is required, leading to smaller cost.

Furthermore, to validate  Algorithm 2  achieves acceleration while guaranteeing DR targeting feasibility, we compare it with the method which solves \eqref{11} directly without firstly judging the feasibility. The computation time under different values of  $\epsilon$ is given in Table \ref{table:time}. This illustrates that the proposed approach can significantly decrease computation time, with the ability to effectively assess feasibility prior to solving MILP problems. In fact, it can achieve a reduction in computation time of over 50\%. This sheds light on practical implementation of real-time DR, which has a hard limit on computation time.

\begin{table}[h]
\caption{Computation time comparison under varying $\epsilon$. }
\begin{center}
\begin{tabular}{c | c c c}
\hline\hline
    & $\epsilon=0.01$ & $\epsilon=0.1$ &$\epsilon=1$
     \\
\hline    \makecell[c]{Proposed} & 232s &
    160s & 118s \\
\hline
\makecell[c]{Original} & 491s & 495s & 435s\\
\hline\hline
\end{tabular}
\end{center}
\label{table:time}
\end{table}

% Concretely, when the number of the targeted nodes is set as 2 or 5, the average LMP cannot be reduced to the reference point, and different values of weight have no impact on the results. When the number of the targeted nodes is set as 10, with less weight imposed on the total demand reduction, the proposed approach can reduce the average LMP to the reference point by reducing more demands. In contrast, with larger weight, the larger demand reduction is punished more. Therefore, the average LMP is higher than the reference point and the corresponding demand reduction is smaller.

\begin{figure}[!ht]
  \centering
  \includegraphics[scale=0.7]{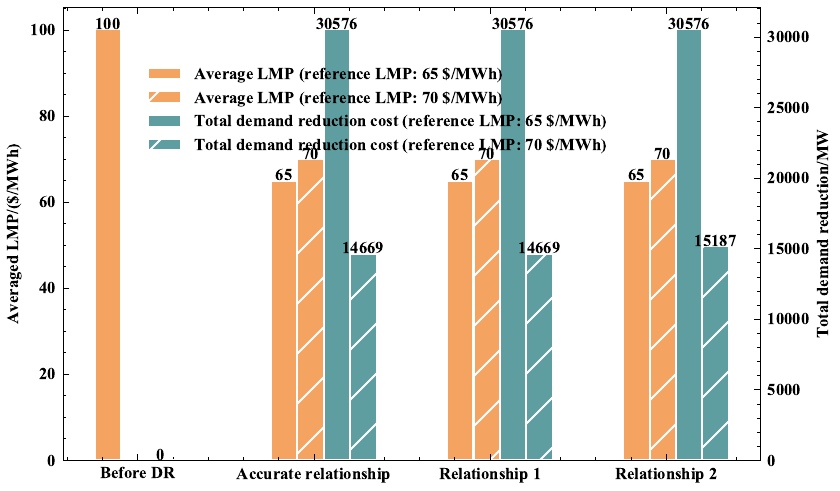}\\
  \caption{The nodal average LMP and total demand reduction cost under the price-demand relationship built by the accurate and inaccurate system parameters.}\label{Fig 8}
\end{figure}

\subsection{Investigation on the Influence of Inaccurate System Parameters}

Due to measurement error and change of operating conditions, recorded system parameters such as line parameters can be different from the actual values. This can pose challenges to practical market operations. To investigate the influence of such inaccurate system parameters on DR targeting results, we consider two inaccurate price-demand relationship built by the inaccurate transmission limits, namely $\bm{\hat{f}}_1=0.9\bm{f}$ and $\bm{\hat{f}}_2=1.1\bm{f}$, where $\bm{f}$ denotes the original line flow limits. And we denote the price-demand relationship built under $\bm{\hat{f}}_1$ and $\bm{\hat{f}}_2$ as relationship 1 and relationship 2 in the following discussion. The results under the relationship based on the accurate parameters and the results under relationship 1 and relationship 2 are shown in Fig. \ref{Fig 8}, where the number of the targeted nodes is set as 5 and the acceptable LMP deviation $\bm{\epsilon}$ is set as 0.01. The reference value of LMP is set as 65 \$/MWh and 70 \$/MWh, respectively.
For the two reference values of LMP, the values of the average LMP after optimization, which relies on the accurate and inaccurate relationships respectively, are the same. The only difference is that the total demand reduction cost under relationship 2 is slightly larger than that under the accurate one, when the reference point is set as 70 \$/MWh. The results show that the impact of the inaccurate parameters on the results is relatively small, and the proposed approach can still achieve the good performance under the price-demand relationship built by the inaccurate system parameters.

\section{Conclusion}

In this study, we introduce an innovative and efficient approach for mitigating price spikes through DR.
% We establish a theoretical relationship between nodal LMPs and nodal demands, alleviating the need for extensive computation and operational efforts associated with solving the ED problem during operations.
Our approach showcases its effectiveness in the context of DR targeting, wherein we proficiently determine the locations for DR implementation and the corresponding quantity of demand reduction at various nodes. Utilizing the derived price-demand relationship, we formulate the DR targeting problem as a rigorous MILP problem. In this framework, demand reduction acts as the ``action knob" for controlling LMPs. Additionally, we propose a solution strategy aimed at enhancing computational efficiency in solving the DR targeting problem.

% In this work, we propose a novel and efficient approach for mitigating price spikes via demand response (DR). The theoretical relationship between the nodal LMPs and the nodal demands is derived, such that the computation and operational efforts of resorting to the ED problem is spared at the operation. We demonstrate its effectiveness on the task of DR targeting, where the locations for implementing DR and the corresponding nodes' demand reduction quantity are effectively determined. With the derived price-demand relationship, the DR targeting problem is formulated as a rigorous MILP problem, where the demand reduction serves as the ``action knob" to control the LMP. A solution strategy is also proposed to improve the computational efficiency for solving the DR targeting problem.

Case studies demonstrate that the conventional heuristic of selecting nodes with the highest LMPs for DR implementation is suboptimal. In contrast, our proposed approach successfully attains the goal of reducing average LMP, a feat unachievable by the heuristic-based method. Also, we validate the performance of our approach across diverse scenarios, including variations in the problem parameters. Furthermore, the experiment shows that the proposed approach can achieve almost the same performance under accurate and inaccurate system parameters. Our approach provides an effective approach for implementing DR. Its application is not limited to mitigating the price spike. Others such as maximizing LSEs' profits through DR can also be explored with the proposed approach. \textcolor{black}{Additionally, the price-demand relationship is derived for ED modeled by a quadratic program. In the future, it is worth to investigating how to derive the relationship for operation problems in other forms.}

% how to target nodes for price reduction under constrained budgets. As the nodes with the highest LMPs are not necessarily the optimal locations for DR, they are ``free-riders" by enjoying the LMP reduction through the demand reduction in other nodes. Thus, future directions about DR mechanism design will include fairness notions regarding customer groups and the uncertainties about both generation and nodal demands.  We will also explore more efficient learning approaches to find the LMP policies.

\section*{Acknowledgement}
Honglin Wen is funded by National Natural Science Foundation of China (grant: 52307119). The authors would like to thank anonymous reviewers for constructive suggestions.

% \bibliographystyle{IEEEtran}
% \bibliography{IEEEabrv,mylib}

\bibliographystyle{elsarticle-harv} 
\bibliography{mylib}

\newpage
\appendix

% \section{Proof that Uniform LMPs across all Nodes Indicate the Absence of Congestion.}\label{Appendix B}

% \begin{proof}
    
%  Here, we show the condition that the nodal LMPs are equal implies no congestion. If LMPs across the nodes are equal, the marginal congestion component in \eqref{2} is a vector of zeros. Since the line flow cannot reach the upper and lower bounds at the same time, it is impossible that $\boldsymbol{\mu_1}^*$ and $\boldsymbol{\mu_2}^*$ equal the same positive value. Therefore, $\boldsymbol{\mu_1}^*$ and $\boldsymbol{\mu_2}^*$ can only equal to  the vector of zeros, which implies that the power flow doesn't reach the lower and upper bounds. Therefore, there is no congestion.

% \end{proof}

\section{Proof of Proposition 1}\label{Appendix A}

\begin{proof}
The Lagrangian of \eqref{6} is given by

\begin{equation}\label{14}
\begin{split}
& L=\frac{1}{2}(\bm{P}^\text{g})^{\top}\bm{Q}(\bm{P}^\text{g})+\bm{q}^{\top}\bm{P}^\text{g}+\sum_{i\in\mathcal{N}}c_i^\text{g}+\\
&\gamma(\bm{1}_{|\mathcal{N}|}^{\top}\bm{P}^\text{g}-\bm{1}_{|\mathcal{N}|}^{\top}\hat{\bm{l}})+\bm{\mu}^{\top}(\bm{A}\bm{P}^\text{g}-\bm{b})+\bm{\psi}^{\top}(\bm{S}\bm{P}^\text{g}-\bm{h})
\end{split}
\end{equation}

Given the optimal solution of primal and dual variables when $\hat{\bm{l}}=\Tilde{\bm{l}}$, namely $\Tilde{\bm{P}^\text{g}},\Tilde{\gamma},\Tilde{\bm{\mu}},\Tilde{\bm{\psi}}$, the KKT conditions for stationarity, primal feasibility, and complementary slackness are

\begin{subequations}\label{15}
\begin{align} 
& \bm{Q}\Tilde{\bm{P}^\text{g}}+\bm{q}+\Tilde{\gamma}\bm{1}_{|\mathcal{N}|}+\bm{A}^\top\Tilde{\bm{\mu}}+\bm{S}^\top\Tilde{\bm{\psi}}=0\\
&\bm{1}_{|\mathcal{N}|}^{\top}\Tilde{\bm{P}^\text{g}}-\bm{1}_{|\mathcal{N}|}^{\top}\hat{\bm{l}}=0\\
&D(\Tilde{\bm{\mu}})(\bm{A}\Tilde{\bm{P}^\text{g}}-\bm{b})=0\\
&D(\Tilde{\bm{\psi}})(\bm{S}\Tilde{\bm{P}^\text{g}}-\bm{h})=0
\end{align}
\end{subequations}
where $D(\cdot)$ creates a diagonal matrix from a vector.

Taking the differentials of these conditions gives the equations
\begin{subequations}\label{16}
\begin{align} 
& (d\bm{Q})\Tilde{\bm{P}^\text{g}}+\bm{Q}(d\bm{P}^\text{g})+d\bm{q}+d\gamma\bm{1}_{|\mathcal{N}|}+(d\bm{A})^\top\Tilde{\bm{\mu}}+\bm{A}^\top(d\bm{\mu})+(d\bm{S})^\top\Tilde{\bm{\psi}}+\bm{S}^\top (d\bm{\psi})=0\\
&\bm{1}_{|\mathcal{N}|}^{\top}d\bm{P}^\text{g}-\bm{1}_{|\mathcal{N}|}^{\top}d\hat{\bm{l}}=0\\
&D(\bm{A}\Tilde{\bm{P}^\text{g}}-\bm{b})d\bm{\mu}+D(\Tilde{\bm{\mu}})[(d\bm{A})\Tilde{\bm{P}^\text{g}}+\bm{A}(d\bm{P}^\text{g})-d\bm{b}]=0\\
&D(\bm{S}\Tilde{\bm{P}^\text{g}}-\bm{h})d\bm{\psi}+D(\Tilde{\bm{\psi}})[(d\bm{S})\Tilde{\bm{P}^\text{g}}+\bm{S}(d\bm{P}^\text{g})-d\bm{h}]=0.
\end{align}
\end{subequations}

We rewrite \eqref{16} into a compact matrix form

\begin{equation}\label{17}
\begin{split}
&\begin{bmatrix}
  \bm{Q} & \bm{1}_{|\mathcal{N}|} & \bm{A}^\top & \bm{S}^\top\\
\bm{1}_{|\mathcal{N}|}^\top & 0 & 0 &0\\
D(\Tilde{\bm{\mu}})\bm{A} & \bm{0} &D(\bm{A}\Tilde{\bm{P}^\text{g}}-\bm{b}) & \bm{0}\\
D(\Tilde{\bm{\psi}})\bm{S} & \bm{0} & \bm{0} & D(\bm{S}\Tilde{\bm{P}^\text{g}}-\bm{h})\\
  \end{bmatrix} 
\cdot
\begin{bmatrix}
d\bm{P}^\text{g}\\
d\gamma\\
d\bm{\mu}\\
d\bm{\psi}
\end{bmatrix}\\
&=\begin{bmatrix}
-d\bm{Q}\Tilde{\bm{P}^\text{g}}-d\bm{q}-(d\bm{A})^\top\Tilde{\bm{\mu}}-(d\bm{S})^\top \Tilde{\bm{\psi}}\\
\bm{1}_{|\mathcal{N}|}^\top d\hat{\bm{l}}\\
-D(\Tilde{\bm{\mu}})(d\bm{A})\Tilde{\bm{P}^\text{g}}+D(\Tilde{\bm{\mu}})d\bm{b}\\
-D(\Tilde{\bm{\psi}})(d\bm{S})\Tilde{\bm{P}^\text{g}}+D(\Tilde{\bm{\psi}})d\bm{h}
\end{bmatrix}
\end{split}
\end{equation}

The coefficient matrix in the left-hand side is the matrix $\bm{M}_0$. Since we wish to compute the Jacobian $\frac{\partial \bm{P}^\text{g}}{\partial \hat{\bm{l}}},\frac{\partial \gamma}{\partial \hat{\bm{l}}},\frac{\partial \bm{\mu}}{\partial \hat{\bm{l}}},\frac{\partial \bm{\psi}}{\partial \hat{\bm{l}}}$, we simply substitute $d\hat{\bm{l}}=\bm{I}$, and set all other differential terms in the right-hand side to zero. Given 
\begin{align*}
d\bm{b}=
\begin{bmatrix}
d\bm{f}+(d\bm{H})\hat{\bm{l}}+\bm{H}d\hat{\bm{l}}\\
d\bm{f}-(d\bm{H})\hat{\bm{l}}-\bm{H}d\hat{\bm{l}}
\end{bmatrix}
\end{align*}
Then the right-hand vector become

\begin{align*}
\begin{bmatrix}
\bm{0}\\
\bm{1}_{|\mathcal{N}|}^\top\\
D(\Tilde{\bm{\mu}})
\begin{bmatrix}
\bm{H}\\
-\bm{H}
\end{bmatrix}\bm{I}\\
\bm{0}
\end{bmatrix},
\end{align*}
which is $\bm{N}_0$. Therefore, we have
\begin{equation}\label{18}
\begin{bmatrix}
    \frac{\partial \bm{P}^\text{g}}{\partial \hat{\bm{l}}}\\
    \frac{\partial \gamma}{\partial \hat{\bm{l}}}\\
    \frac{\partial \bm{\mu}}{\partial \hat{\bm{l}}}\\
    \frac{\partial \bm{\psi}}{\partial \hat{\bm{l}}}
\end{bmatrix}
=\bm{M}_0^{-1}\bm{N}_0
\end{equation}

With the \textbf{Theorem 1} and the Jacobian derived in \eqref{18}, the calculation of $\bm{P}^\text{g}_m(\hat{\bm{l}}),\gamma_m(\hat{\bm{l}})$, $\bm{\mu}_m(\hat{\bm{l}}),\bm{\psi}_m(\hat{\bm{l}})$, in the subregion of $\hat{\bm{l}}=\Tilde{\bm{l}}$ is the affine function defined in \eqref{7}, which completes the proof.

\end{proof}

\section{The formulation of comparison approach.}\label{Appendix B}
The comparison approach solves an optimal power flow problem, with the objective of minimizing the generation cost and the demand reduction cost. Let $\bm{r}=[r_i]_{i \in \mathcal{N}}$ be the vector of demand reduction, where $r_i=0$ if the node $i$ is not chosen as DR location. The problem is formulated as,
\begin{subequations}
\begin{align} \mathop{\min}_{\bm{P}^\text{g},\bm{r}} \quad & \frac{1}{2}\bm{P}^{\text{g}\top}\bm{Q}\bm{P}^{\text{g}}+\bm{q}^\top\bm{P}^{\text{g}}+\sum_{i\in\mathcal{N}}c_i^\text{g}+\bm{\tau}^\top\bm{r}\\ 
    \text{s.t.} \quad & \sum_{i=1}^{|\mathcal{N}|}P_i^\text{g}=\sum_{i=1}^{|\mathcal{N}|}l_i-r_i:\gamma
    \\ 
    &
    -\bm{f}\leq \bm{H}(\bm{P}^\text{g}-\bm{l}+\bm{r})\leq\bm{f}:\bm{\mu}_1,\bm{\mu}_2\\
    & \bm{P}^{\text{min}} \leq \bm{P}^\text{g} \leq \bm{P}^{\text{max}}: 
    \bm{\psi}_1,\bm{\psi}_2\\
    & \bm{0} \leq \bm{r} \leq \bar{\bm{x}}
\end{align}
\end{subequations}

After solving the above problem, the LMP is obtained via \eqref{2}.

%\addtolength{\textheight}{rs in the-12cm}   % This command serves to balance the column lengths

%%%%%%%%%%%%%%%%%%%%%%%%%%%%%%%%%%%%%%%%%%%%%%%%%%%%%%%%%%%%%%%%%%%%%%%%%%%%%%%%

% \bibliographystyle{IEEEtran}

% \bibliography{IEEEabrv,mylib}

\end{document}